\documentclass[sigconf]{acmart}
\AtBeginDocument{%
  }

\setcopyright{acmlicensed}
\copyrightyear{2026}
\acmYear{2026}
\acmDOI{XXXXXXX.XXXXXXX}
\acmConference[ICSE SEET '26]{IEEE/ACM ICSE Software Engineering Education and Training}{April 14--16,
  2026}{Rio de Janeiro, BR}
\acmISBN{978-1-4503-XXXX-X/2018/06}
\usepackage{xcolor}
\usepackage{tcolorbox}
\usepackage[inkscapelatex=false]{svg}
\newcommand{\qql}[1]{\textcolor{black}{{#1}}}

\begin{document}

\title{On the Role and Impact of GenAI Tools in Software Engineering Education}


\author{Qiaolin Qin}
\email{qiaolin.qin@polymtl.ca}
\affiliation{%
  \institution{Polytechnique Montreal}
  \city{Montreal}
  \country{Canada}
}

\author{Ronnie de Souza Santos}
\email{ronnie.desouzasantos@ucalgary.ca }
\affiliation{%
  \institution{University of Calgary}
  \city{Calgary}
  \country{Canada}
}

\author{Rodrigo Spinola}
\email{spinolaro@vcu.edu}
\affiliation{%
  \institution{Virginia Commonwealth University}
  \city{Richmond}
  \country{United States}
}

\renewcommand{\shortauthors}{Qin et al.}

\begin{abstract}
\textit{Context}. The rise of generative AI (GenAI) tools like ChatGPT and GitHub Copilot has transformed how software is learned and written. In software engineering (SE) education, these tools offer new opportunities for support, but also raise concerns about over-reliance, ethical use, and impacts on learning. 
\textit{Objective}. This study investigates how undergraduate SE students use GenAI tools, focusing on the benefits, challenges, ethical concerns, and instructional expectations that shape their experiences. 
\textit{Method}. We conducted a survey with 130 undergraduate students from two universities. The survey combined structured Likert-scale items and open-ended questions to investigate five dimensions: usage context, perceived benefits, challenges, ethical and instructional perceptions. 
\textit{Results}. Students most often use GenAI for incremental learning and advanced implementation, reporting benefits such as brainstorming support and confidence-building. At the same time, they face challenges including unclear rationales and difficulty adapting outputs. Students highlight ethical concerns around fairness and misconduct, and call for clearer instructional guidance.
\textit{Conclusion}. GenAI is reshaping SE education in nuanced ways. Our findings underscore the need for scaffolding, ethical policies, and adaptive instructional strategies to ensure that GenAI supports equitable and effective learning.
\end{abstract}



\maketitle

\section{Introduction}
\label{sec:introduction}

The adoption of generative AI (GenAI) tools, including ChatGPT, GitHub Copilot, and Gemini, is reshaping how software is developed, studied, and maintained~\cite{bommasani2021opportunities,denny2024computing,vaithilingam2022expectation,nguyen2022empirical}. Within educational contexts, such tools are increasingly embedded in students’ practices for information seeking, code generation, debugging, and problem conceptualization~\cite{becker2023programming,prather2023robots,ziegler2022productivity}. This development signals a departure from long-established conventions in computing education, where achievement has historically depended on individual understanding, manual work, and instructor guidance. Instead, students today can access context-aware suggestions, natural language explanations, and even full-fledged implementations with a few keystrokes.

This new paradigm creates both opportunities and problems. On one hand, GenAI tools can reduce entry barriers, accelerate task completion, and foster experimentation~\cite{liang2024usability,white2023prompt}. On the other hand, they introduce concerns around academic integrity, over-reliance, and the erosion of critical thinking and problem-solving skills~\cite{ko2024calculators,prather2023robots}. These tensions are particularly pronounced in software engineering (SE) education, where learning outcomes emphasize not just coding, but understanding trade-offs, system design, collaboration, and professional ethics. As a result, educators and institutions are now pressed to reevaluate pedagogical approaches, assessment methods, and policy frameworks to keep pace with GenAI's influence on the student experience~\cite{lau2023ban,wang2023adapting}.

A recent qualitative study investigated how software engineering students interact with GenAI tools~\cite{choudhuri2024frontline}. Based on interviews with 16 upper-level students, the study proposed a framework consisting of four usage contexts and five categories of challenges. The contexts framework—L1 (initial learning), L2 (incremental learning), I1 (initial implementation), and I2 (advanced implementation)—describe the timing and manner of student engagement with GenAI in coursework. This segmentation appeared to facilitate the identification of distinct benefits, such as support for brainstorming and syntax, as well as difficulties, including prompt formulation and misalignment with learning preferences.

Their challenge framework describes recurring difficulties students face: unclear understanding of GenAI and its use (C1), difficulty in communicating needs or context (C2), misalignment with student learning processes or preferences (C3), lack of rationale in GenAI responses (C4), and difficulty in appropriately applying GenAI outputs (C5). These challenges were found to be rooted in intrinsic flaws and limitations of current GenAI systems, such as hallucinations, lack of scaffolding, and reasoning gaps, and were associated with cognitive and affective consequences ranging from confusion and frustration to reduced trust and learning efficacy.


While this usage and challenge framework has provided a useful perspective on the role of GenAI in software engineering education~\cite{choudhuri2024frontline}, it was developed from a small, single-institution sample of 16 students. Such a scope constrains the extent to which its insights can be generalized. In addition, the study did not examine how ethical and instructional considerations affect students’ experiences with GenAI, dimensions that appear increasingly relevant as these tools are adopted in classrooms and begin to shape learning practices.

To build on this foundation, our study surveys 130 undergraduate SE students from two large universities. We investigate when, how, and why students use GenAI tools, as well as the benefits and challenges they encounter. In this sense, we expand the research on this topic to explicitly consider ethical concerns and instructional expectations, dimensions that are critical for guiding the responsible integration of GenAI in SE education.

Guided by this motivation, we pose the following overarching research question:
\textbf{How do undergraduate SE students experience and perceive generative AI tools in their coursework, in terms of usage, benefits, challenges, ethical concerns, and instructional expectations?}
This overarching question is unpacked through five specific research questions (RQ1–RQ5), which investigate distinct dimensions of usage contexts, perceived benefits, challenges, ethical perceptions, and instructional expectations.


Overall, our study makes three main contributions:
\begin{enumerate}
    \item We extend the previously published usage and challenge framework~\cite{choudhuri2024frontline} with data from 130 software engineering students across two institutions (RQ1–RQ3).
    \item We provide insights into students’ ethical perceptions and instructional expectations regarding GenAI in software engineering education, indicating the importance of clear policies and adaptable course design (RQ4–RQ5).
\end{enumerate}

In addition to this introduction, the paper is organized into six further sections. Section~\ref{sec:related_work} provides background on GenAI in computing education and reviews prior work, including the Usage/Challenge framework. Section~\ref{sec:methodology} presents the study design, including the research questions, survey instrument, participant demographics, and data analysis plan. Section~\ref{sec:results} reports the findings of the survey across all five research questions. Section~\ref{sec:discussion} interprets the results in light of previous research and highlights implications. Section~\ref{sec:threats_to_validity} discusses threats to validity and study limitations. Finally, Section~\ref{sec:conclusion} concludes the paper and outlines directions for future work. Our survey data are publicly available at: \hyperref[]{https://anonymous.4open.science/r/On-the-Role-and-Impact-of-GenAI-Tools-in-Software-Engineering-Education-E980/README.md}.

\section{Background and Related Work}
\label{sec:related_work}

The growing presence of GenAI tools in computing education has sparked a wave of empirical and theoretical investigations. Researchers have begun to explore not only the capabilities and limitations of tools like ChatGPT and GitHub Copilot but also their pedagogical implications, ethical ramifications, and effects on student learning. This section situates our work within this emerging body of research. We first review studies examining GenAI use in SE education broadly, then focus on the dual usage/challenge framework developed by Choudhuri et al., which serves as the foundation for our study. 

\subsection{GenAI in Software Engineering Education}
\label{sec:genai_in_see}

The integration of GenAI tools into SE education has sparked both enthusiasm and concern. These tools can assist in code generation, debugging, and documentation, potentially enhancing learning efficiency and productivity \cite{vaithilingam2022expectation}. However, they also raise challenges related to code plagiarism, learning dependency, and uneven skill acquisition \cite{finnieansley2022robots,li2023impact, cotton2023chatgpt, rudolph2023chatgpt, lund2023chatgpt, kasneci2023chatgpt}. The impacts are to be discussed from both students' and instructors' sides. 

\textbf{While GenAI tools are widely applied by students, their usage patterns are different, which may lead to uneven outcomes on knowledge acquisition.} Denny et al. \cite{denny2023student} analyzed large-scale data from programming courses and found significant variation in how students used AI tools, from passive reference checks to direct code completion. \textbf{Further, empirical studies have shown that while students are eager to adopt GenAI tools, their use often leads to shallow engagement with course content.} For instance, Vaithilingam et al. \cite{vaithilingam2022expectation} found that novice programmers using GitHub Copilot perceived a tension between achieving task completion and gaining deeper learning. Similarly, Finnie-Ansley et al. \cite{finnieansley2022robots} reported that students using code-generating tools might bypass problem-solving efforts and rely too heavily on AI-generated answers. Li et al. \cite{li2023impact} highlighted that while students appreciated ChatGPT’s support, its uncritical use could result in conceptual misunderstandings and overconfidence in incorrect output.

\textbf{On the other hand, recent studies revealed that GenAI tools are reshaping the role of instructors and assessments.}  
Ghosh et al. \cite{ghosh2023genai} explored how instructors in computing courses have adapted assignments and grading practices in response to the widespread use of GenAI tools, noting a shift toward more design-based and open-ended tasks. \textbf{Instructors also face uncertainty about what constitutes fair and pedagogically sound use of GenAI.} Frick et al. \cite{frick2023instructors} reported that many instructors feel underprepared to develop policies, design AI-resilient assignments, or evaluate student learning in the presence of AI-generated content. Moreover, the ethical and legal implications of using GenAI in coursework remain under debate, especially around authorship, attribution, and data privacy \cite{nguyen2023legal}.

As GenAI tools become more deeply embedded in software development workflows, understanding their educational role becomes increasingly important. These tools are not just productivity enhancers, they shape how students conceptualize problem-solving and evaluate the correctness of solutions. Our work builds on this context by extending the framework introduced by Choudhuri et al. \cite{choudhuri2024frontline}, which captures students’ experiences, struggles, and strategies with GenAI tools in SE coursework.

\subsection{The Usage/Challenge Framework}
\label{sec:study_framework}

Choudhuri et al.~\cite{choudhuri2024frontline} conducted a qualitative study based on interviews with 16 SE students to investigate how GenAI tools are being used to support learning and implementation tasks in SE education, resulting in two interrelated conceptual \qql{usage context and challenge frameworks} that characterize \textit{when}, \textit{how}, and \textit{why} students use GenAI, and the challenges and consequences of that use.

The first part (i.e., the usage framework) identifies four primary contexts of use across the SE \textbf{learning and implementation} pipeline: initial learning (L1), incremental learning (L2), initial implementation (I1), and advanced implementation (I2). In the learning context, L1 corresponds to situations where students engage with GenAI to grasp SE concepts from scratch, often lacking prior knowledge. L2 reflects instances where students build upon existing understanding, using GenAI for clarification, revision, or practice. In the implementation context, I1 captures the use of GenAI for bootstrapping tasks such as setting up frameworks, generating boilerplate code, or brainstorming ideas. I2 encompasses more advanced uses, including code optimization, integration, and refinement within ongoing projects.

The authors found that students experienced the most benefits during L2 and I1 phases. In these contexts, GenAI was perceived as helpful in providing structured guidance, personalized explanations, and initial direction, reducing friction when reviewing concepts or starting a coding task. In contrast, challenges were most prominent during L1 and I2 phases. These phases involved either low student familiarity with the topic or high task complexity, revealing important mismatches between student needs and GenAI capabilities.

To explain these challenges, the second part of the framework (i.e., the challenge framework) outlines five categories of recurring issues (C1-C5). C1 refers to students’ unclear understanding of GenAI’s limitations and ethical use. C2 captures their difficulty in communicating needs or context effectively, such as crafting prompts or providing sufficient background. C3 involves misalignments between GenAI responses and student preferences, processes, or learning styles. C4 reflects the lack of clear rationales in GenAI output, which hinders deeper comprehension. Finally, C5 concerns students’ struggles to verify, adapt, or meaningfully integrate GenAI responses into their work.

These challenges were not isolated but often cascaded into one another, with C2 (communication issues) and C4 (lack of rationale) particularly affecting both learning and task performance. Importantly, the authors found that these issues had broader consequences beyond task outcomes, affecting student confidence, willingness to adopt GenAI tools, and perceptions of their own learning. 

By mapping the contexts in which students use GenAI tools and the challenges they encounter, Choudhuri et al.’s framework offers a foundational lens for understanding GenAI use in SE education. Building on this foundation, our study extends the investigation to a broader and more diverse student population, while also foregrounding the ethical and instructional concerns that arise when GenAI becomes embedded in learning practices.
 

\section{Method}
\label{sec:methodology}
In this study, we adopt a survey methodology as the empirical approach to collect evidence from 130 software engineering students across two institutions. Surveys are a common empirical method in SE, used to identify patterns across populations. In this research, we followed established guidelines for survey research in software engineering \cite{kitchenham2008personal, ralph2020empirical} to ensure methodological rigor and to account for the specific characteristics of SE contexts.

\subsection{Research Questions}
\label{sec:research_questions}

This study aims to provide a comprehensive understanding of how undergraduate SE students interact with GenAI tools in educational contexts. To achieve this, we formulated five research questions (RQs), each addressing a key dimension (i.e., usage contexts, benefits, challenges, ethical implications, and instructional implications) of student experience with GenAI tools. These questions were derived through a synthesis of prior qualitative findings~\cite{choudhuri2024frontline}, instructional challenges, and emerging concerns in computing education \cite{ghosh2023genai, frick2023instructors, nguyen2023legal}.

\begin{itemize}
  \item[\textbf{RQ1:}] \textit{In what contexts do SE students use GenAI tools, and how frequently?}

  This question seeks to quantify the frequency and purpose of GenAI tool usage across distinct learning and implementation scenarios. It builds upon the usage framework~\cite{choudhuri2024frontline}, enabling us to explore whether students engage with GenAI differently when acquiring new concepts versus implementing code.

  \item[\textbf{RQ2:}] \textit{What are the perceived benefits of using GenAI tools in SE learning and project work?}

  Prior studies suggest that GenAI tools support brainstorming, comprehension, and productivity~\cite{choudhuri2024frontline}. This question assesses which benefits students most frequently report and how those benefits relate to their learning practices and confidence levels.

  \item[\textbf{RQ3:}] \textit{What challenges do students face when using GenAI tools in SE coursework, and how do these relate to known categories of difficulty?}

  This question operationalizes the challenge framework~\cite{choudhuri2024frontline} and quantifies the prevalence of issues such as prompt ambiguity, learning misalignment, and over-reliance.

  \item[\textbf{RQ4:}] \textit{How do students perceive the ethical implications of GenAI usage in SE education, including issues related to fairness, academic integrity, and responsible use?}

  This question addresses the emerging ethical climate around GenAI adoption. It explores students’ beliefs regarding misconduct, fairness, and the social acceptability of using GenAI tools in the classroom.

  \item[\textbf{RQ5:}] \textit{How do students perceive the instructional implications of GenAI usage in SE education?}

  Beyond ethics, this question considers how GenAI influences instructional design. It investigates student preferences regarding guidance, training, and course adaptation in light of AI’s growing presence.


\end{itemize}

Answering these research questions is essential for understanding the evolving role of GenAI tools in SE education. As these tools become increasingly integrated into the student experience, empirical insights into how, why, and with what effects they are used will inform evidence-based pedagogical strategies. By capturing not only usage patterns but also student-reported benefits, challenges, ethical concerns, and instructional expectations, this study aims to equip educators, administrators, and policymakers with the knowledge needed to foster responsible, equitable, and effective use of GenAI in SE curricula.

\subsection{Survey Design}
\label{sec:survey_design}

\begin{table*} [!t] 
\centering
\caption{Survey Questions (simplified) and their corresponding titles. Closed-end questions for student background collection are labeled as SC (Single Choice) or MC (Multiple Choice); the remaining closed-end questions are in single choice likert-scale styles.}
\label{tab:survey-questions}
\resizebox{\textwidth}{!}{%
\begin{tabular} {cllc}
\hline
\textbf{Research Questions} & \multicolumn{1}{c}{\textbf{Question No.-Title}} & \multicolumn{1}{c}{\textbf{Question}} & \textbf{Type} \\
\hline
\textbf{Background} & \textbf{Q1-Study Year} & What year are you in the program? & Closed(SC)\\
\textbf{Collection} & \textbf{Q2-SE Coursework}  & Which SE courses have you completed? & Closed(MC)\\
 & \textbf{Q3-Tool Exposure} & Which GenAI tools have you used for coursework? & Closed(MC)\\
 & \textbf{Q4-Usage Frequency} & How often do you use GenAI tools for coursework? & Closed(SC)\\
 & \textbf{Q5-Programming Skill} & What is your self-rated programming skill level? & Closed(SC)\\
\hline

\textbf{RQ1} &\textbf{Q6-Basic Concept Learning}&I use GenAI tools to gain a basic understanding of new SE concepts. & Likert Scale\\
\textbf{Contexts of} &\textbf{Q7-Concept Refinement}& I use GenAI to refine or clarify SE concepts I already partially understand. & Likert Scale\\
\textbf{Use} &\textbf{Q8-Start Coding Help}& I use GenAI to get started with my coding assignments (e.g., boilerplate code). & Likert Scale\\
 &\textbf{Q9-Code Optimization}& I use GenAI to optimize, refactor, or debug production-level code in my projects. & Likert Scale\\
\hline
\textbf{RQ2} &\textbf{Q10-Success Story}& Describe a situation when GenAI helped you succeed in an SE task. & Open ended\\
\textbf{Perceived} &\textbf{Q11-Brainstorming Support}& GenAI helps me brainstorm and outline implementation steps.& Likert Scale\\
\textbf{Benefits} &\textbf{Q12-Concept Application}& GenAI helps me understand how to apply SE concepts in practice.& Likert Scale\\
 &\textbf{Q13-Clear Examples}& GenAI provides clear examples that improve my understanding.& Likert Scale\\
 &\textbf{Q14-Confidence Boost}& GenAI boosts my confidence when starting a coding task.& Likert Scale\\

\hline

\textbf{RQ3} &\textbf{Q15-Confusion Example}& Describe a time when GenAI caused confusion or frustration. & Open ended\\
\textbf{Perceived} &\textbf{Q16-Misconduct Uncertainty}& I am unsure when using GenAI constitutes academic misconduct. & Likert Scale\\
\textbf{Challenges} &\textbf{Q17-Prompt Difficulty}& I find it hard to phrase questions in a way that GenAI understands. & Likert Scale\\
 &\textbf{Q18-Learning Style Mismatch}& GenAI explanations do not match the way I learn best. & Likert Scale\\
 &\textbf{Q19-Missing Rationales}& GenAI rarely explains why a solution works, even if it seems correct. & Likert Scale\\
 &\textbf{Q20-Response Adaptation}& I struggle to adapt GenAI responses to the specific needs of my assignments. & Likert Scale\\
 &\textbf{Q21-Assignment Frustration}& Using GenAI has left me frustrated or confused during assignments. & Likert Scale\\
 &\textbf{Q22-Wrong Concepts}& I have had to ``unlearn'' incorrect concepts provided by GenAI. & Likert Scale\\

\hline

\textbf{RQ4} &\textbf{Q23-Misconduct Definition}& In your opinion, when does AI usage become a form of academic misconduct? & Open ended\\
\textbf{Ethical} &\textbf{Q24-Instructor Policies}&My instructors have clearly communicated policies about the acceptable use of AI tools. & Likert Scale\\
\textbf{Perspectives} &\textbf{Q25-Acknowledgment Ethics}&I consider using AI tools in assignments to be ethically acceptable when properly acknowledged. & Likert Scale\\
 &\textbf{Q26-Unfair Advantage Concern}&I am concerned that some students may have an unfair advantage by using AI tools. & Likert Scale\\
 &\textbf{Q27-Comprehension Requirement}&I believe that using GenAI without understanding the generated output is ethically problematic. & Likert Scale\\
 &\textbf{Q28-Peer Pressure}&I feel pressure to use AI tools because others do, even if I’m unsure whether it’s allowed. & Likert Scale\\
 &\textbf{Q29-Support Policies}&I would support clear institutional policies on the ethical boundaries of AI tool use in coursework. & Likert Scale\\

\hline
\textbf{RQ5} &\textbf{Q30-Integration Suggestion}& Describe one way AI tools could be better integrated into SE instruction or assignments. & Open ended\\
\textbf{Instructional} &\textbf{Q31-Teaching Changes}&I believe AI tools like ChatGPT and Copilot are changing the way SE should be taught. & Likert Scale\\
\textbf{Expectations}&\textbf{Q32-Guidance Need}&I would like my instructors to provide more guidance on how to use AI tools responsibly. & Likert Scale\\
&\textbf{Q33-Long-Term Risks}&I think relying on AI tools for assignments can negatively affect my learning in the long term. & Likert Scale\\
&\textbf{Q34-Evaluation Confidence}&I feel confident in evaluating whether AI-generated code is correct and appropriate. & Likert Scale\\
&\textbf{Q35-Training Benefit}&I would benefit from formal training on how to effectively use AI tools in SE tasks. & Likert Scale\\
&\textbf{Q36-Assignment Adaptation}&I believe assignment design should adapt to the presence of AI tools. & Likert Scale\\
&\textbf{Q37-Adequate Instruction}&I have received adequate in-class instruction or resources to help me evaluate AI-generated output. & Likert Scale\\ \hline
\end{tabular}%
}
\end{table*}

The survey instrument was developed to address RQ1-RQ5, with content organized into six sections reflecting the constructs of interest: 
(1) student background, 
(2) contexts of GenAI use, 
(3) perceived benefits, 
(4) perceived challenges, 
(5) ethical perspectives, and
(6) instructional expectations. 
The mapping between survey questions and research questions is detailed in Table~\ref{tab:survey-questions}, ensuring full coverage of the study’s conceptual framework.

Each section of the instrument was grounded in prior work and refined to match the operational definitions of our constructs:

\begin{itemize}
    \item \textbf{Background (Q1--Q5)} captured demographic and experiential factors, including academic standing, completed SE courses, prior GenAI use, frequency of use, and self-rated programming skill. These items contextualize patterns observed across other sections. 
    \item \textbf{Contexts of Use (Q6--Q9)} measured the frequency of GenAI use across the four usage contexts described by Choudhuri et al.~\cite{choudhuri2024frontline}, using Likert-scale items to quantify how often students engaged in each context. This section directly informs RQ1.
    \item \textbf{Perceived Benefits (Q10--Q14)} included items reflecting benefits previously identified in the literature (e.g., brainstorming support, code comprehension, productivity gains) and refined through the usage/challenge framework~\cite{choudhuri2024frontline}. This section addresses RQ2.
    \item \textbf{Perceived Challenges (Q15--Q22)} operationalized the challenge categories from Choudhuri et al.~\cite{choudhuri2024frontline}, with items targeting issues such as unclear tool understanding, prompt formulation difficulty, misalignment with learning style, lack of rationale in outputs, and difficulty applying results. These items inform RQ3.
    \item \textbf{Ethical Perspectives (Q23--Q29)} gauged students’ level of agreement with statements on academic integrity, fairness, authorship, and acceptable use scenarios. Prior studies have highlighted ethical risks such as plagiarism, misuse, and equity concerns when GenAI tools are adopted in educational contexts \cite{lund2023chatgpt, kasneci2023chatgpt}. By operationalizing these themes in our survey, we directly address RQ4 on how students perceive the ethical implications of GenAI in SE education.
    \item \textbf{Instructional Expectations (Q30--Q37)} captured student preferences regarding instructor guidance, training, and curriculum adjustments to ensure responsible and effective GenAI integration. Research has underscored the pressing need for clear instructional frameworks and scaffolding to help students navigate GenAI use productively and ethically \cite{cotton2023chatgpt, rudolph2023chatgpt}. Accordingly, these questions respond to RQ5 by examining how students envision the role of faculty and curriculum design in supporting sustainable GenAI adoption.
\end{itemize}

The survey used a mix of \textit{multiple-choice}, \textit{5-point Likert-scale}, and \textit{open-ended} questions. Likert items ranged from ``Strongly Disagree'' (1) to ``Strongly Agree'' (5), to capture the intensity of students’ agreement with statements across different categories. Open-ended questions invited students to describe positive or negative experiences with GenAI in SE coursework and provide suggestions for effective instructional integration, enabling thematic insights beyond structured items.

To ensure \textbf{content validity}, items were developed through iterative refinement. An initial draft was based on the Choudhuri et al. usage/challenge framework~\cite{choudhuri2024frontline}. The draft was reviewed by two researchers for clarity and alignment with the RQs, then piloted with a small group \qql{(i.e., 5 participants)} of upper-level SE students to check for interpretability and completion time. Minor revisions were made to wording and ordering to reduce ambiguity and cognitive load. The final instrument ensures \textbf{traceability between items, constructs, and RQs} (Table~\ref{tab:survey-questions}), enabling targeted analysis for each research dimension while also allowing cross-sectional and correlational examination across dimensions.

\subsection{Data Collection}
\label{sec:data_collection}

Data was collected through an online survey administered to undergraduate students enrolled in upper-level SE courses at Virginia Commonwealth University (VCU) and the University of Calgary. These students are expected to have intermediate to advanced programming experience and prior exposure to GenAI tools in academic contexts.

The survey was distributed during weeks 2–4 of the semester to ensure early engagement while allowing sufficient course exposure to contextualize responses. Students were invited to participate during class sessions and provided with a secure link to the online questionnaire. Prior to participation, students were presented with an informed consent statement outlining the purpose, procedures, and voluntary nature of the study. 

To encourage participation, students were offered an optional incentive, such as entry into a raffle for a \$50 Amazon gift card. The survey took around 10 minutes to complete and was accessible via mobile and desktop devices. No personally identifiable information was collected, ensuring the anonymity and confidentiality of all participants.

\subsection{Data Analysis}
\label{sec:data_analysis}

We applied both quantitative and qualitative analysis techniques to examine the responses collected through the survey instrument. We leverage single-choice and multiple-choice questions to collect participants' background information. The remaining closed-ended questions are in Likert-scale format and were quantitatively analyzed using descriptive statistics (e.g., frequencies, percentages, and measures of central tendency) to summarize students' perceptions, usage patterns, and expectations regarding GenAI tools in SE education. 

The open-ended questions aim to gain a more in-depth context of students' experience in using GenAI for educational purposes. Responses to these questions are used to qualitatively explain and enrich our findings from the closed-ended questions. 


This mixed-methods approach enabled us to triangulate findings across data types, enhancing the richness and depth of our interpretation. Quantitative results provided a broad overview of student's opinion trends, while qualitative insights helped contextualize and explain these patterns in students' own words.

\section{Results}
\label{sec:results}

\begin{figure*}[h!]
  \center
  \includegraphics[keepaspectratio=1,width=0.9\textwidth]{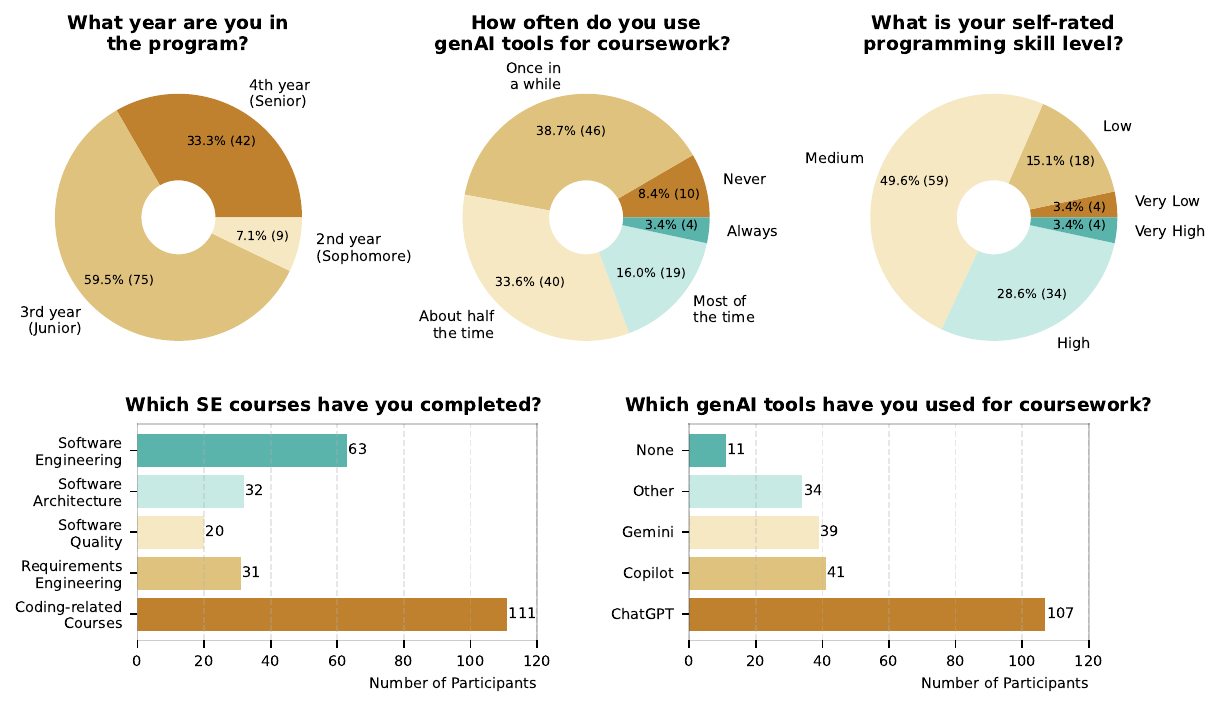}
  \caption{The participants' background information collected in Q1-Q5. }
  \label{fig:background}
\end{figure*}

We collected 130 student responses from two universities; 86 students completed the mandatory questions, while 44 students partially completed the survey. \qql{Since every response reflects participants' opinions on the studied aspects, we still consider the partially completed responses during our analysis. In other words, our findings from each survey question are based on the collected answers instead of the whole participant distribution.} Participants' background information collected from Q1-Q5 is summarized in Fig.~\ref{fig:background}: the responses of single-choice questions are visualized with pie charts, and the responses for multiple-choice questions are visualized with bar plots. Most of the students participating in our survey are in their third or fourth year of undergraduate study (Q1), have taken at least one SE-related course (Q2), and have medium-to-high-level programming skills (Q5). Apart from 11 students, the remaining participants for Q4 reported that they have used at least one of the GenAI tools for coursework. According to the responses for Q3, the most popular tool among students is ChatGPT. 

\begin{figure*}[h!]
  \center
  \includegraphics[keepaspectratio=1,width=\textwidth]{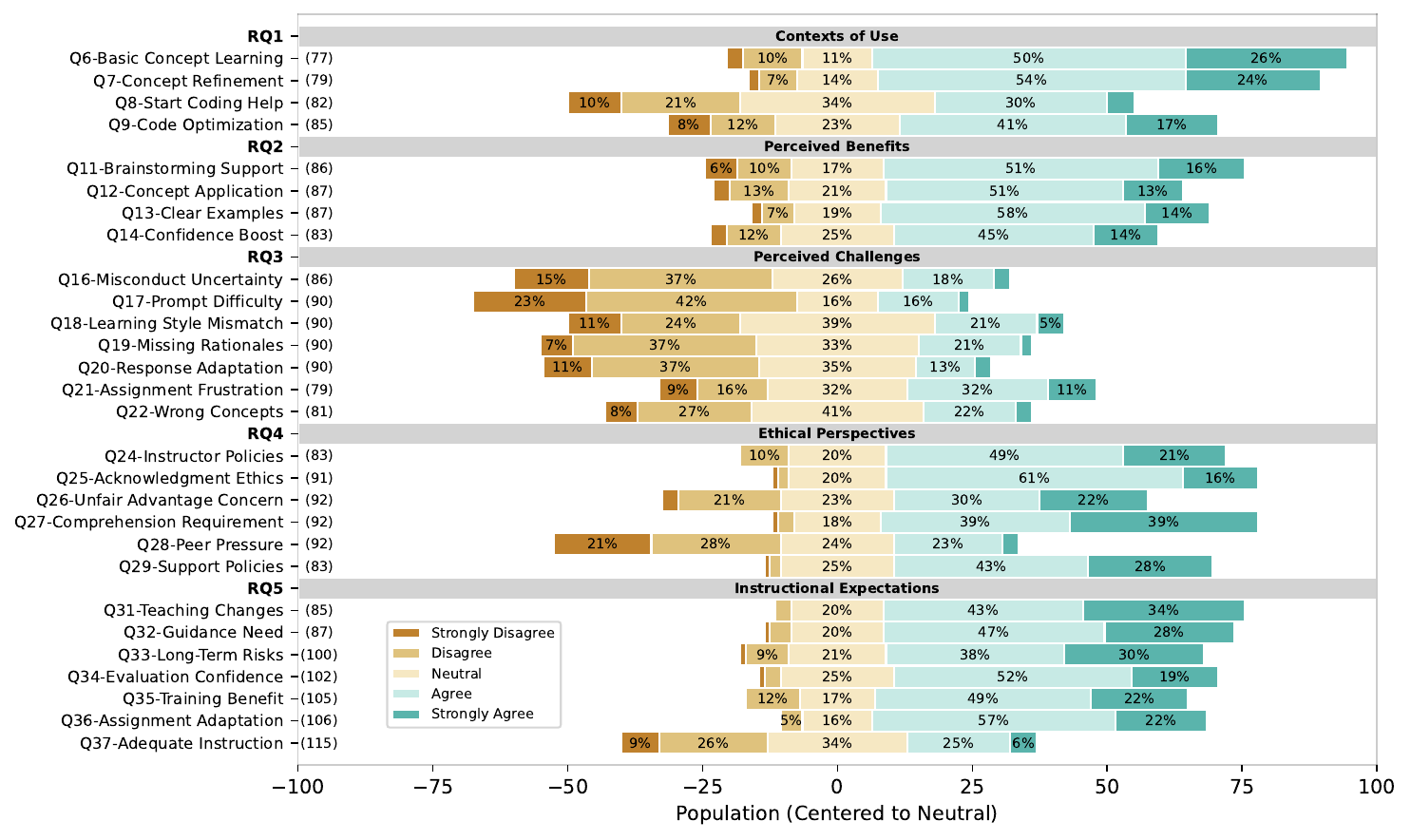}
  \caption{Likert chart of our closed-ended questions results, the number of responses is labeled next to each question title. The questions are segmented according to the five sections introduced in Sec.~\ref{sec:survey_design}: context of use, perceived benefits, perceived challenges, ethical perspectives, and instructional expectations. Minor responses (i.e., response ratios of less than 5\%) are hidden in the caption for improved visualizations. }
  \label{fig:likert_chart}
\end{figure*}

The survey results for the remaining 28 closed-ended questions are illustrated in Fig.~\ref{fig:likert_chart}. In the following subsections, we answer each research question by quantitatively analyzing the Likert scale results and cite students' responses to qualitatively explain these findings. 

\subsection*{RQ1: In what contexts do SE students use GenAI tools, and how frequently?}
\label{sec:results_rq1}

\textbf{Majority of the students leverage GenAI to learn SE concepts.} More than 75\% of participants agreed that they use GenAI to gain a basic and deeper understanding of SE concepts (L1 and L2) in Q6 and Q7, and the responses' mean values for both questions are around 3.9, showing a strong inclination to agreement. The usage of GenAI for concept learning is commonly mentioned in Q10. Some participants mentioned that GenAI can sometimes provide explanations more accessible than course materials: ``\textit{I was able to use GenAI to explain fundamental operating system concepts, specifically memory fragmentation. I found that the explanation provided by the AI was better than the notes provided to me by the class} (\#186093856)''.  

\textbf{Students use GenAI more in enhancing their code than in starting projects.} The successful usage of GenAI for code implementation is also widely described in Q10. Many mentioned that GenAI tools serve as a strong aid in improving code quality by fixing bugs, creating test cases, or adding standardized documentation: ``\textit{(AI) words it in a generally accepted manner, and I don’t have to wonder whether my words translate to another} (\#185384161)''. The general trend of using GenAI for code improvement (I2) is supported by the results of Q9, as 58\% of the participants hold an agreement opinion. In comparison, the opinions on using GenAI in starting coding for a project (I1) are more diverse: 31\% of the participants tend not to use GenAI to start their project, 35\% hold the opposite opinion, and 34\% of the participants are unsure (i.e., neutral) about this usage scenario. The mean response value of Q8 is 3.0, suggesting no usage pattern can be summarized. This outcome is implicitly reflected in Q10: few participants shared their experiences in using GenAI to start coding a project from scratch. 

An interesting discovery in Q8 is that more students have a strongly disagreeing opinion (10\%) about starting their GenAI assignments than strongly agree (<5\%). The reason for such a result is related to both bad experiences and ethical concerns, which are discussed in RQ3 and RQ4.

\subsection*{RQ2: What are the perceived benefits of using GenAI tools in SE learning and project work?}
\label{sec:results_rq2}

\textbf{Students agree that GenAI tools are beneficial in enhancing their concept comprehension.} 64\% of participants agreed that GenAI can help them in understanding SE concepts in practice (Q12), and a higher ratio of participants (72\%) pointed out that GenAI helps provide examples to aid understanding (Q13). This benefit is also widely mentioned in Q10, given that GenAI tools can better customize the explanations in comparison to existing forum answers. For instance, a participant mentioned that GenAI ``\textit{helped me explain some Java concepts when search engines failed} (\#185383952)''. 

\textbf{Students find GenAI beneficial for task planning instead of starting coding.} It is worth noticing that students share a high agreement rate in Q14, while no trend of agreement/disagreement can be summarized from Q8. The seemingly conflicting results suggest that students still use GenAI at the beginning of a coding task (I1), but not for coding purposes. Indeed, corresponding to the 67\% agreement rate in Q11, participants mentioned many scenarios of GenAI tools aiding brainstorming implementation plans in Q10, such as ``\textit{to give me a plan and how to start a plan} (\#185383875)'' and ``\textit{helping me discover libraries to import and make a task easier} (\#185383968)''.

\subsection*{RQ3: What challenges do students face when using GenAI tools in SE coursework, and how do these relate to known categories of difficulty?}
\label{sec:results_rq3}

\textbf{\qql{Few students face the ethical challenges (C1) during GenAI usage}.} In educational scenarios such as finishing course assignments, students from the two institutes are instructed on whether GenAI usage is acceptable by course lecturers. Therefore, a relatively low portion of participants (22\%) expressed ethical concerns (C1) in Q16. 

\textbf{Students are generally able to communicate with GenAI effectively, but \qql{they sometimes encounter difficulties during communication (C2)}.} The result of Q17 received a mean response value of 2.3, showing that most of the students (67\%) did not encounter severe communication problems with GenAI (C2). However, 86 students reported that GenAI caused confusion or frustration during their usage in Q15, mentioning that ``\textit{GenAI does not understand my prompt or its implementation} (\#185384160)''. 

\textbf{\qql{Students report that GenAI has response quality problems (C3-C5) due to model hallucination or lack of domain knowledge}.} C3-C5 focus on different aspects under GenAI response quality (i.e., misalignment in preferences, absence of response rationale, and difficulty in response implementation). While most of the participants reported that they did not experience critical problems in response preference, logic, and implementability, we noticed that more than 1/3 of the participants voted ``neutral'' for Q18-Q20 and Q22. The neutral opinion suggests that students encountered response quality problems during usage, although not highly frequently. The problems are often due to the models' hallucinations or lack of domain knowledge. Students mentioned GenAI has ``\textit{limited scope of understanding, GPT-4 would sometimes hallucinate code between prompts} (\#185412493)'', and ``\textit{certain coursework uses very niche technologies that GenAI’s suggestions do not work at all with and just cause more confusion} (\#186099660)''. 

As a result, they have to verify the credibility of responses and put extra effort into looking for the correct solutions: ``\textit{the answers are never guaranteed to be correct, and that causes confusion during studying} (\#185384191)''. Although the faults can be easily identified sometimes ``\textit{when I know the answer and I know that it is wrong} (\#185383875)'', the concrete risk of using GenAI roots in learning new concepts: a student mentioned that the tools were ``\textit{giving me wrong examples of things I'm unfamiliar with} (\#185383978)''. 

\textbf{Concerning the communication difficulties and response quality problems, students are usually unsatisfied when using GenAI for their study.} According to Q21, only 25\% of participants are generally satisfied with their learning experience using GenAI. Students widely expressed their frustrations in Q15, reporting that GenAI provides them ``\textit{incorrect answers over and over} (\#185384131)'', and are sometimes not helpful because the tools were ``\textit{generating code or ideas that are not clear on how it was written or the process} (\#185384359)''.

\subsection*{RQ4: How do students perceive the ethical implications of GenAI usage in SE education, including issues related to fairness, academic integrity, and responsible use?}
\label{sec:results_rq4}

\textbf{Institutional instructions can help students to understand the acceptability of GenAI usage.} In Q24, 70\% of participants from the two institutions reported that their lecturers instructed the acceptability of GenAI usage for their courses. This result corresponds to our findings in Q16. With guidance from instructors, most participants have a clear understanding of the appropriateness of GenAI usage. 

With adequate in-class ethical instructions, 59\% of participants voted against using GenAI when they're unsure about the acceptability of usage, even if other students are doing so (Q28). However, 41\% of participants are neutral or inclined to use GenAI due to the uncertainty of usage allowance. Further, 52\% of participants are concerned about some students obtaining an unfair advantage by using GenAI tools (Q26). The potential peer pressure and unfair advantage concerns call for institutional ethical regulations. Indeed, the mean response value for Q29 reached 3.9, highlighting students' vast support for clear institutional policies on GenAI usage. 

\textbf{Students are concerned about the authorship and credibility in GenAI usage.} In Q25, 77\% of participants consider acknowledging the usage of GenAI tools as an ethical practice. Nonetheless, even with potential crediting to GenAI, the direct use of responses generated by these tools is still concerning. 78\% of the participants believe that it is unacceptable to use the generated output without comprehension (Q27), as this action poses questions about the originality \qql{(i.e., whether the answer reflects the student's knowledge)} and credibility \qql{(i.e., whether the answer is correct and reliable)} of the assignment, and leads to unfair justifications in assignment scoring or competitions for prizes. 

Based on the above concerns, many participants defined academic misconduct involving GenAI tools in Q23 as ``\textit{not doing any of the work and just copy-paste your code and submit without even checking} (\#185383975), ``\textit{when it’s used to cheat on tests} (\#185384247)'', and more generally, ``\textit{when it becomes your catalyst for success rather than a tool to help you} (\#185383852)''.

\subsection*{RQ5: How do students perceive the instructional implications of GenAI usage in SE education?}
\label{sec:results_rq5}

\textbf{Students believe that GenAI tools are changing the education pattern in SE.} Corresponding to GenAI's perceived benefits on gaining a better understanding of SE concepts, more than 70\% of participants agreed that GenAI tools such as ChatGPT and Copilot are changing the teaching pattern in SE (Q31). Given GenAI's powerful inference capability, students can use GenAI to ``\textit{write pseudocode/algorithms and help understand a problem you’re working on} (\#185384366)''. They also agreed that guidance on GenAI usage from instructors can aid their SE study (Q32 and Q35). In Q30, participants proposed that instructors' guidance could include ``\textit{teaching students how to use it to help their learning instead of using it as a crutch} (\#185384025)'' and ``\textit{giving clearer instructions or a better framework about how to use AI for coursework} (\#185474764)''. 

Further, 79\% of the participants agreed that assignments should change due to the presence of GenAI (Q36). First, they mentioned that, as a powerful tool, GenAI ``\textit{should be allowed to be used on assignments} (\#186093532)''. Second, since basic function coding is easily replaceable by GenAI tools, participants proposed having assignments ``\textit{based on how well you can use it} (\#185383852)''. Given that GenAI ``\textit{allows students to create more complex programs and assignments} (\#185384247)'', a probable change in assignment could be ``\textit{allowing students to use it and expect the assignment or task to be harder or have more layers} (\#186104948)''. 

\textbf{The usage of GenAI in completing assignments should be regulated.} As previously discussed in RQ3, GenAI may provide misleading answers that teach students incorrect information. Although participants agreed that assignments should embrace the existence of GenAI, the negative impact brought by relying on GenAI tools for finishing assignments is highly concerning (Q33). Since the usage of GenAI tools is an inevitable trend, the evaluation of AI content in coursework is needed to control students' dependency on GenAI and reduce the previously mentioned ethical problems. As a type of institutional policy, 71\% of participants support the evaluation of AI content in their project code (Q34). 

However, the mean value of 2.9 for Q37 suggests that participants felt that the monitoring of GenAI contents is still inadequate. Instructors should improve their approaches in GenAI evaluation. For example, as proposed by one of the participants, instructors could ``\textit{ask more questions about our code so we can show we understand it, and if we don’t, that’s how you know we’re taking advantage of AI} (\#185384161)''.


\section{Discussion}
\label{sec:discussion}

Our study contributes to the growing body of work examining the role of GenAI in software engineering education by assessing the usage/challenge framework of Choudhuri et al.~\cite{choudhuri2024frontline} (RQ1-RQ3) with a larger and more diverse student population. Moreover, this work also brings forward discussions on the ethical and instructional implications of GenAI use (RQ4 and RQ5), highlighting students’ perceptions of fairness, integrity, and the evolving role of teaching practices. In this section, we interpret our findings across the five research questions, reflect on their implications for pedagogy and policy, and identify avenues for future research.  

\subsection{Patterns of Use and the Usage/Challenge Framework}
Our results reinforce the four-phase framework proposed by Choudhuri et al.~\cite{choudhuri2024frontline}, but they also surface important nuances. Students reported frequent reliance on GenAI for incremental learning (L2) and advanced implementation (I2), aligning with prior reports that AI assistance is most effective when students already possess baseline knowledge or when refining existing code~\cite{denny2024computing,finnieansley2022robots}. By contrast, initial learning (L1) and initial implementation (I1) produced more mixed results: while some students found GenAI useful in lowering barriers to entry, others distrusted the quality of output in these phases. 

These findings suggest that the framework captures \textit{when} students use GenAI but does not fully explain \textit{why} some contexts are more beneficial than others. Student confidence, prior exposure to AI tools, and institutional context mediate these differences. Pedagogically, instructors should encourage GenAI use in contexts where it is most supportive (incremental refinement and advanced debugging), while providing safeguards and explicit scaffolding in riskier contexts (initial learning and first implementations).  

\subsection{Perceived Benefits and Pedagogical Opportunities}
Students emphasized several benefits of GenAI, including brainstorming support, access to diverse examples, and accelerated comprehension of SE concepts. These align with prior work showing that students often turn to AI tools for inspiration and rapid feedback~\cite{denny2023student,liang2024usability}. Many students also highlighted that GenAI-generated explanations felt more tailored and accessible than traditional resources such as textbooks or Q\&A forums.  

This creates new pedagogical opportunities. For example, instructors can integrate GenAI into classroom exercises where students critically compare AI-generated examples with instructor-provided or peer-created solutions. Similarly, GenAI’s brainstorming utility can be harnessed in collaborative design sessions, allowing students to generate multiple approaches that are then critiqued and refined. Rather than banning GenAI, educators should treat it as a resource for active learning, where its generative capabilities serve as a starting point for deeper analysis and reflection.  

\subsection{Challenges and Limitations of GenAI in SE Learning}
Despite the benefits, students reported substantial challenges, particularly in line with categories C3–C5 of the prior framework: misalignment with learning preferences, lack of rationale, and difficulty adapting AI responses. Students frequently described frustrations when AI outputs appeared correct but lacked explanations, making it harder to understand \textit{why} a solution worked. This resonates with findings from Li et al.~\cite{li2023impact} and Vaithilingam et al.~\cite{vaithilingam2022expectation}.  

Our survey suggests that while communication difficulties (C2) were less dominant, response quality and reliability remain major concerns. Students often had to spend additional time verifying GenAI outputs, which could either reinforce learning or cause discouragement. The risk is particularly acute for novices, who may lack the expertise to identify subtle but consequential errors.  

The implication for SE education is that curricula must explicitly teach verification, testing, and critical assessment of GenAI outputs. Assignments should require students to demonstrate how they validated AI results. For instance, students could annotate AI-generated code with justifications, comparisons to alternatives, or tests verifying correctness. Such strategies transform AI’s limitations into opportunities for cultivating rigorous engineering habits.  

\subsection{Ethical Perceptions and Academic Integrity}
The ethical dimension of GenAI use is complex. While most students supported acknowledgment of GenAI use and rejected “copy-paste” misuse as academic misconduct, many expressed concerns about fairness, authorship, and unregulated advantage. These concerns mirror prior work highlighting instructors’ uncertainty about defining ethical boundaries~\cite{frick2023instructors,nguyen2023legal}.  

Our results suggest that ethical clarity is uneven: some students reported clear guidance from instructors, while others faced ambiguity and peer pressure to adopt tools even when unsure of their permissibility. This highlights the need for institutional-level policies that are consistent, transparent, and co-developed with both faculty and students.  

Beyond policy, ethics education itself must evolve. Rather than framing GenAI as simply “permitted” or “forbidden,” curricula should include discussions of responsible use, attribution practices, and the professional norms students will encounter in industry. Embedding these topics into SE ethics courses, capstones, or project-based assignments could prepare students to navigate real-world contexts where GenAI is ubiquitous but rarely regulated.  

\subsection{Instructional Implications and Course Design}
A clear majority of students agreed that GenAI is reshaping SE education and should be explicitly addressed in teaching. They expressed strong support for formal training on responsible use and for redesigning assignments to reflect the realities of AI-assisted coding. These perspectives align with broader recommendations to shift toward open-ended, design-oriented tasks~\cite{lau2023ban,wang2023adapting}\qql{, such as asking students to design solutions for a given type of tasks}.  

Our results suggest several strategies for course design:  
\begin{itemize}
    \item \textbf{Assignment redesign:} Move from syntax-heavy tasks to assignments requiring evaluation, integration, and justification of GenAI outputs.  
    \item \textbf{Scaffolding and training:} Offer workshops or in-class modules on prompt engineering, error detection, and AI evaluation practices.  
    \item \textbf{Assessment adjustments:} Incorporate oral exams, code reviews, or reflective essays where students explain \textit{how} they used GenAI, not just \textit{what} it produced.  
    \item \textbf{Adaptive complexity:} Increase assignment complexity when GenAI is allowed, shifting learning goals from code writing to design, testing, and analysis.  
\end{itemize}

Such adaptations help ensure that GenAI augments rather than replaces student learning. However, they also demand significant instructor effort, including the creation of AI-resilient tasks and the management of evolving policies.  




\subsection{Key Takeaways}
The usage/challenge framework continues to serve as a valid lens for understanding the role of GenAI in software engineering education, though the outcomes are strongly mediated by students’ backgrounds and their level of confidence. Students reported clear benefits, such as support for brainstorming, access to personalized examples, and a sense of confidence-building; these advantages can be effectively leveraged within active and collaborative learning environments. At the same time, students faced recurring challenges that included hallucinations, the absence of clear rationales in tool outputs, and difficulties adapting generic responses to their specific needs. These issues point to the necessity of explicit training in critical evaluation skills.

Ethical concerns further highlight the importance of developing clearer, co-designed institutional policies and embedding responsible AI use within curricula. Instructional design also needs to evolve in response to GenAI, for example by raising the complexity of assignments, incorporating structured scaffolding, and emphasizing reflective practice. Finally, equity considerations remain central: differentiated support and institutional measures are required to ensure that GenAI serves to close rather than widen learning gaps.

\section{Threats to Validity}
\label{sec:threats_to_validity}

We reflect on the potential threats to the validity of our study using the four established categories: construct, internal, external, and conclusion validity~\cite{wohlin2012experimentation}. 

\textbf{Construct Validity.} Construct validity concerns whether the instrument effectively measures the intended concepts. Our survey was designed to capture students’ expectations, perceptions, and intended use of GenAI tools in SE education. To enhance construct validity, we built on validated conceptual frameworks from prior work, particularly the usage/challenge framework by Choudhuri et al.~\cite{choudhuri2024frontline}, and conducted a pilot with students to refine the survey questions for clarity and relevance.

However, since the survey was administered at the beginning of the semester, students may not yet have engaged with GenAI tools in the context of the specific course in which the survey was distributed. As a result, some responses, particularly those related to expected instructional support or course, may be based on assumptions. Nonetheless, many students likely drew on prior experiences with GenAI tools gained through other SE related courses, internships, or independent use. Therefore, while some responses may reflect expectations, others are likely informed by concrete past experiences, especially regarding general benefits, challenges, and ethical concerns associated with GenAI in SE contexts.

\textbf{Internal Validity.} Internal validity relates to the soundness of causal inferences within the study. Although our study is exploratory and non-experimental (and thus not aiming to establish causality), we acknowledge several internal threats. First, self-selection bias remains a concern: students who chose to respond may differ in motivation, background, or familiarity with GenAI compared to those who did not. This could influence observed patterns.

Notably, because the survey was administered at the very beginning of the semester, it avoids recency bias, i.e., responses are unlikely to be influenced by isolated recent events or course-specific frustrations. 

\textbf{External Validity.} External validity refers to the extent to which findings can be generalized beyond the study sample. While our sample includes 130 students from two institutions, they were all enrolled in upper-level SE courses. As such, our results may not reflect the views of students from other disciplines, educational levels (e.g., freshmen), or institutions with different levels of GenAI tool adoption.

However, the cross-institutional design of this study enhances its external validity. By collecting data from two geographically and institutionally distinct universities, we introduce variation in instructional context, institutional policy, and student demographics, which allows us to capture a more diverse set of perspectives and practices. This diversity strengthens the generalizability of our findings, making them more likely to reflect broader patterns in SE education across different academic environments. Still, caution should be taken when applying our results to different educational systems or cultural settings. 

\textbf{Conclusion Validity.} Conclusion validity concerns whether the inferences drawn from the collected data are reasonable and supported by the analysis. To strengthen conclusion validity, we employed a mixed-methods approach that combined descriptive statistics with responses to the open-ended questions. This triangulation allowed us to cross-validate key trends and contextualize quantitative results with student narratives, reducing the likelihood of spurious conclusions.
Nevertheless, several threats remain. First, our data is based on self-reported perceptions and behaviors, which are inherently subjective and may not always align with actual tool usage. Although open-ended responses provided richer insight, the accuracy of conclusions relies on students’ introspective abilities and willingness to share candidly. Second, since our analysis focuses on descriptive and correlational patterns rather than causal inference, observed relationships should be interpreted with caution.

Another consideration is the temporal framing of the data collection. The survey was administered early in the semester, which may limit the depth of students’ experiences with GenAI in that particular course. While many participants drew on prior experiences across other courses or contexts, their responses regarding challenges, instructional expectations, or ethical tensions may evolve over time. Longitudinal follow-ups would be necessary to fully assess how GenAI usage and its impacts change as students engage more deeply with course content and AI tools.

Finally, the usage/challenge framework used to structure our analysis, while empirically grounded, was adapted from a prior qualitative study and applied in a new context and at a broader scale. Our operationalization of its constructs through survey items involved some degree of interpretation. Although we aligned items with the original constructs and conducted a pilot to refine them, measurement error or construct drift cannot be ruled out entirely.

\section{Concluding Remarks}
\label{sec:conclusion}
This study makes two main contributions. First, it provides an empirical validation and extension of a usage and challenge framework previously proposed in the literature~\cite{choudhuri2024frontline}, drawing on data from more than 130 software engineering students across two institutions (RQ1–RQ3). Second, it examines students’ ethical perceptions and instructional expectations regarding GenAI in software engineering education, indicating the importance of clear policies and adaptive course design (RQ4–RQ5).

Our findings suggest that while GenAI tools are widely adopted for incremental learning and advanced implementation, their role in initial learning and project initiation remains contested. Students reported benefits such as brainstorming support, personalized examples, and confidence-building, but also noted recurring challenges, including unclear rationales, hallucinations, and difficulties in adapting generated content. These patterns indicate the dual nature of GenAI as both a facilitator of learning and a potential source of misconceptions if used without critical reflection.

Beyond technical considerations, our results point to the need for explicit ethical guidance and adaptive instructional strategies. Students expressed interest in clearer policies and in assignments that incorporate GenAI in ways that promote, rather than undermine, critical thinking. Therefore, our research suggests that software engineering educators, administrators, and policymakers should engage proactively with GenAI integration. Promising strategies include scaffolding responsible use, redesigning assessments, and ensuring fairness in both access and evaluation.

Future research could build on this foundation by conducting longitudinal studies to track how student practices and perceptions evolve, investigating cross-cultural differences in adoption, and designing interventions that help students critically evaluate AI-generated content. Such directions appear essential to promote not only efficiency, but also deep learning, equity, and professional responsibility in software engineering education.

\bibliographystyle{ACM-Reference-Format}
\bibliography{references}
\end{document}